# Quantum control and manipulations with stationary three-color lights

S. A. Moiseev[1,2][*] and B. S. Ham[1]

[1]The Graduate School of Information and Communications, Inha University, Incheon 402-751 S. Korea
[2]Kazan Physical-Technical Institute of Russian Academy of Sciences, Kazan 420029 Russia

E-mails: samoi@yandex.ru; bham@inha.ac.kr

A dynamic quantum control of three-color lights in an optically dense medium is presented. We discuss how effectively to stop traveling three-color light pulses in the medium by using three control laser fields at near resonant transitions satisfying electromagnetically induced transparency. This opens a door to the quantum coherent control of multiple traveling light pulses for quantum memory and quantum switching, which are essential components in multi-party quantum optical communications.

## I. INTRODUCTION

Quantum coherent control has drawn much attention in the nonlinear quantum optics especially in a very weak light region [1-3]. Of very attractive possibilities for using such quantum coherent controls are adiabatic interactions of the lights with the resonant media [4]. A big progress in the control of a weak light has been obtained recently using electromagnetically induced transparency (EIT) [5] and successfully applied to quantum nonlinear optics for a giant Kerr nonlinearity [6], photon switching [7,8], and quantum memory [9,10]. EIT is a quantum optical phenomenon, where matter-field interactions modify the refractive index of the medium to be transparent even to a resonant light [11-13]. This type of absorption spectrum control is a basis of quantum manipulation of a group velocity of a light pulse, where correlation between atomic coherence and photonic components plays a major role [14,15].

Recently, a standing-wave-grating based stationary light technique based on EIT using a hot atomic vapor has been demonstrated [16]. In general, the interaction time of a traveling light with an optical medium is independent of the group velocity of the light, whereas a stationary light can enormously increase the interaction time. This property looks attractive for nonlinear quantum optics for especially a spatially limited optical medium, where enhanced nonlinear effects is a direct result of both EIT-induced absorption free and the lengthened photon-atom interaction time. The manipulations with such stationary lights contribute to the enhancement of intrinsically weak nonlinear photon-atom interactions [17] and generation of an entangled state of photons [18]. To effectively enhance the weak-field limited nonlinearity, strong spatial confinement and/or precise spectral/temporal manipulations of the light are essential [17]. For this, several different methods using stationary light have been proposed [19-21].

In this paper we present a stationary three-color light phenomenon which can be realized using either cold atomic gases or condensed media. We theoretically analyze the three-color stationary light for frequency conversion and photonic trapping, where the photonic trapping is relevant to the quantum memory. Using a quantum field propagating through an optical medium, we present quantum manipulation of three-color stationary light using strong classical laser fields. This one color quantum field based three color stationary light is analyzed for effective quantum wavelength conversion. We also discuss potential applications of the three-color stationary light for nonlinear quantum optics.

Figure 1 shows an energy level diagram of the present three-color stationary light. Three weak quantum fields $E_{+,1}$, $E_{+,2}$ and $E_{-,1}$ are near resonant to the four-level medium derived by the three additional control laser fields. The two-control fields with Rabi frequencies $\Omega_{+,1}$ and $\Omega_{+,2}$ copropagate along with the fields $E_{+,1}$ and $E_{+,2}$, whereas the field $E_{-,1}$ propagates backward along with the control laser field $\Omega_-$ (with Rabi frequency $\Omega_-$). We assume the frequency difference is $|\omega_+ - \omega_-| \gg \delta\omega_p$ ($\delta\omega_p$ is a spectral width of each light pulse), where $\omega_{+,1}$ and $\omega_{-,1}$ are frequencies of $E_+$ and $E_-$, the $\omega_{31}$ is a resonant frequency of $|1\rangle$ - $|3\rangle$, and the frequency difference satisfies $|\omega_+ - \omega_-| \gg \delta\omega_p$ ($\delta\omega_p$ is a spectral width of the slow light pulses). Here, the propagation direction $\mathbf{k}_-$ of the field $E_-$ is determined by the phase matching with Bragg condition: $\mathbf{k}_- = \mathbf{k}_+ - \mathbf{K}_+ + \mathbf{K}_-$. Unlike intensity-grating-based standing light [16], the present scheme can utilize quantum wavelength conversion using a traveling monochromatic quantum light, in which all three-spectral components are absorption free.

## II. THEOTY OF THE THREE-COLOR STATIONARY LIGHT

For a theoretical analysis of three-color stationary light based on a double−Λ scheme in Fig. 1, we introduce a quantum field $E_\sigma = \sqrt{\hbar\omega_\sigma/(2\varepsilon_o V)} A_\sigma e^{-i(\omega_\sigma t - k_\sigma z)} + H.C.$, where σ stands for the two forward field $\sigma = +,1; +,2$ and one backward field $\sigma = -,1$ which are traveling in ±z direction, respectively, where $A_\sigma$ is slowly varying field operator; we assume that the quantization volume V=1. In the interaction picture the Hamiltonian of the quantum fields with atoms driven by the control laser fields $\Omega_{\pm,m}(t) = \Omega_{\pm,om}(t) e^{i\varphi_{\pm,m}(t)}$ ( $\Omega_{\pm,om}(t)$ and $\varphi_{\pm,m}(t)$ are respectively slowly varying Rabi frequencies on the atomic transition $|1\rangle - |3\rangle$ and phases of the fields):

$$H = -\hbar \sum_{j=1} \{g_1[\hat{A}_{+,1}(t,z_j)e^{-i(\Delta_{+,1}t-k_{+,1}z_j)} + \hat{A}_{-,1}(t,z_j)e^{-i(\Delta_{-,1}t+k_{-,1}z_j)}]P_{31}^j + g_2\hat{A}_{+,2}(t,z_j)e^{-i(\Delta_{+,2}t-k_{+,2}z_j)}P_{41}^j\}$$

$$-\hbar \sum_{j=1}\{[\Omega_{+,1}(t)e^{-i(\Delta_{+,1}t-K_{+,1}z_j)} + \Omega_{-,1}e^{-i(\Delta_{-}t+K_{-}z_j)}]P_{32}^j + \Omega_{+,2}(t)e^{-i(\Delta_{+,2}t-K_{+,2}z_j)}P_{42}^j\} + H.C., \quad (1)$$

where $P_{nm}^j = (P_{mn}^j)^+$ are the atomic operators, $g_\sigma = \wp_\sigma\sqrt{\omega_\sigma/(2\varepsilon_o\hbar V)}$ is a coupling constant of photons with atoms, $\wp_\sigma$ is a dipole moment for each transition [1], $\Delta_{\pm,1} = \omega_{\pm,1} - \omega_{31}$, $k_{\pm,1} = \omega_{\pm,1}^p/c$, $K_\pm = \omega_\pm^c/c$. Using Eq. (1) we derive Heisenberg equations for the atomic operators $P_{mn}^j$ in the limit of weak probe fields $\hat{A}_\pm$ and adding the decay constants $\gamma$ and $\gamma_2$ for the atomic transitions |1>-|3> and |1>-|2>:

$$\tfrac{\partial}{\partial t}P_{14}^j = -\gamma_4 P_{14}^j + ig\hat{A}_{+,2}(t,z_j)e^{-i(\Delta_{+,2}t-k_{+,2}z_j)} + i\Omega_{+,2}e^{-i(\Delta_{+,2}t-K_{+,2}z_j)}P_{12}^j + \hat{f}_4(t,z_j), \quad (2a)$$

$$\tfrac{\partial}{\partial t}P_{13}^j = -\gamma_3 P_{13}^j + ig\{\hat{A}_+(t,z_j)e^{-i(\Delta_+ t-k_+ z_j)} + \hat{A}_-(t,z_j)e^{-i(\Delta_- t+k_- z_j)}\}$$
$$+ i\{\Omega_+ e^{-i(\Delta_+ t-K_+ z_j)} + \Omega_- e^{-i(\Delta_- t+K_- z_j)}\}P_{12}^j + \hat{f}_3(t,z_j), \quad (2b)$$

$$\tfrac{\partial}{\partial t}P_{12}^j = -\gamma_2 P_{12}^j + i\{\Omega_{+,1}^* e^{i(\Delta_{+,1}t-K_{+,1}z_j)} + \Omega_-^* e^{i(\Delta_- t+K_- z_j)}\}P_{13}^j$$
$$+ i\Omega_{+,2}^* e^{i(\Delta_{+,2}t-K_{+,2}z_j)}P_{14}^j + \hat{f}_2(t,z_j), \quad (2c)$$

where $\hat{f}_{2,3,4}(t,z_j)$ are the Langevin forces associated with the relaxation processes [1]. In Eqs. (2), however, we ignore the influences of the forces in the adiabatic limit [22], because it is more convenient due to both the lengthening of the interaction time in the stationary light and suppression of the fluctuation forces in the limit $\Delta_{\pm,1} \gg \gamma_3$ and $\Delta_{+,2} \gg \gamma_4$. Assuming $|\Delta_{\pm,1}|\delta t \gg 1$ and $|\Delta_{+,2}|\delta t \gg 1$ ($\delta t$ is a temporal duration of the probe pulse $E_{+,1}$) and substituting Eqs. (2) and (3) by $P_{13} = P_+ e^{-i(\Delta_{+1}t-k_{+,1}z_j)} + P_- e^{-i(\Delta_{-,1}t+k_{-,1}z_j)}$ and $P_{14} = \tilde{P}_{14}e^{-i(\Delta_{+,2}t-k_{+,2}z_j)}$, the following simplified system equations for are obtained for slowly varying amplitudes in both time and space:

$$\tfrac{\partial}{\partial t}\tilde{P}_{14} = -\gamma_{+2}\tilde{P}_{14} + ig_2 A_{+,2} + i\Omega_{+,2}e^{-ik_o z}\tilde{P}_{12}, \quad (3a)$$

$$\tfrac{\partial}{\partial t}P_+ = -\gamma_{+1}P_+ + ig_1 A_{+,1} + i\Omega_{+,1}e^{-ik_o z}P_{12}, \quad (3b)$$

$$\tfrac{\partial}{\partial t}P_- = -\gamma_{-1}P_- + ig_1 A_{-,1} + i\Omega_{-,1}e^{ik_o z}P_{12}, \quad (3c)$$

$$\tfrac{\partial}{\partial t}P_{12} = -\gamma_2 P_{12} + i\{\Omega_+^* e^{ik_o z}P_+ + \Omega_-^* e^{-ik_o z}P_-\} + i\Omega_{+,2}^* e^{i(\Delta_{+,2}t-K_{+,2}z_j)}P_{14}, \quad (3d)$$

where $k_o = \omega_{21}/c$, $\gamma_{\pm,1} = \gamma_3 - i\Delta_\pm$, $\gamma_{+,2} = \gamma_4 - i\Delta_+$. For a long enough temporal duration of the probe field pulse with $|\gamma_\pm|\delta t \gg 1$, a typical slow light condition under electromagnetically induced transparency is satisfied [1,5]. Under this condition, we find the following coherence operators $P_{\pm} \cong i\gamma_{\pm,3}^{-1}\{g_1 A_{\pm,1} + \Omega_{\pm,1} e^{\mp ik_o z} P_{12}\}$ and $\widetilde{P}_{14} \cong i\gamma_{+,4}^{-1}\{g_2 A_{+,2} + \Omega_{+,2} e^{-ik_o z} P_{12}\}$. For long lived spin coherence:

$$\tfrac{\partial}{\partial t} P_{12} = -\mu_s P_{12} - (F_{+,1} + F_{+,2} + F_{-,1}), \tag{4}$$

where $\mu_s(t) = (\gamma_2 + \gamma_{+,1}^{-1}\Omega_{+,o1}^2(t) + \gamma_{+,2}^{-1}\Omega_{+,o2}^2(t) + \gamma_{-,1}^{-1}\Omega_{-,o1}^2(t))$, $F_{\pm,m} = \Omega_{\pm,om}^2(t)\Psi_{\pm,m}(t,z)/(\gamma_{\pm,m}\sqrt{N})$ (N is atomic density), we obtain a new field operators $\Psi_{\pm,1} = e^{\pm ik_o z}\sqrt{N} g_1 A_{\pm,1}/\Omega_{\pm,1}(t)$ and $\Psi_{+,2} = e^{ik_o z}\sqrt{N} g_2 A_{+,2}/\Omega_{+,2}(t)$ [10]. Thus, Eq. (3) become:

$$(\tfrac{\partial}{c\partial t} + \tfrac{\partial}{\partial z})A_{+,2} = i(Ng_2/c)\widetilde{P}_{+,2}, \tag{5a}$$

$$(\tfrac{\partial}{c\partial t} + \tfrac{\partial}{\partial z})A_{+,1} = i(Ng_1/c)P_{+,1}, \tag{5b}$$

$$(\tfrac{\partial}{c\partial t} - \tfrac{\partial}{\partial z})A_{-,1} = i(Ng_1/c)P_{-,1}. \tag{5c}$$

We now study slow light evolutions with switching processes by the control lasers under the typical adiabatic condition for the slow light propagations, $\gamma \delta t >> 1$ and $\delta t |\mu_s(t)| >> 1$, so that $\widetilde{P}_{12} \cong \widetilde{P}_{12}^{(0)} + \widetilde{P}_{12}^{(1)}$, where $\widetilde{P}_{12}^{(0)} = -(F_{+,1} + F_{+,2} + F_{-,1})/\mu_s$ $\widetilde{P}_{12}^{(1)} = \tfrac{\partial}{\mu_s \partial t}\{(F_{+,1} + F_{+,2} + F_{-,1})/\mu_s\}$. The first term $\widetilde{P}_{12}^{(0)}$ determines the main contribution to the atomic coherence between the levels |1> and |2> in the adiabatic limit. After substitution of $\widetilde{P}_{12}$ into Eqs. (3) and (4) we obtain the following coupled wave equations under the slow-light propagation $v_{\pm,1} = c\Omega_{\pm,1}^2/Ng_1^2 << c$ and $v_{+,2} = c\Omega_{+,2}^2/Ng_2^2 << c$:

$$\xi_{+,2}^{-1}(\tfrac{\partial}{\partial z} - ik_o)\Psi_{+,2} = -\{(\widetilde{\gamma}_2 + \alpha_{+,1} + \alpha_{-,1})\Psi_{+,2} - (\alpha_{+,1}\Psi_{+,1} + \alpha_{-,1}\Psi_{-,1}) + \sqrt{N}\widetilde{P}_{12}^{(1)}\}, \tag{6a}$$

$$\xi_{+,1}^{-1}(\tfrac{\partial}{\partial z} - ik_o)\Psi_{+,1} = -\{(\widetilde{\gamma}_2 + \alpha_{+,2} + \alpha_{-,1})\Psi_{+,1} - (\alpha_{+,2}\Psi_{+,2} + \alpha_{-,1}\Psi_{-,1}) + \sqrt{N}\widetilde{P}_{12}^{(1)}\}, \tag{6b}$$

$$\xi_{-,1}^{-1}(\tfrac{\partial}{\partial z} + ik_o)\Psi_{-,1} = \{(\widetilde{\gamma}_2 + \alpha_{+,2} + \alpha_{+,1})\Psi_{-,1} - (\alpha_{+,2}\Psi_{+,2} + \alpha_{+,1}\Psi_{+,1}) + \sqrt{N}\widetilde{P}_{12}^{(1)}\}, \tag{6c}$$

where $\xi_{+,2} = (\gamma_4/\gamma_{+,2})\xi_{1-4}$, $\xi_{\pm,1} = (\gamma_3/\gamma_{\pm,1})\xi_{1-3}$, $\widetilde{\gamma}_2 = \gamma_2/\mu_s$, $\alpha_{\pm,1}(t) = \Omega_{\pm,o1}^2(t)/(\mu_s(t)\gamma_{\pm,1})$ and $\alpha_{+,2}(t) = \Omega_{+,o2}^2(t)/(\mu_s(t)\gamma_{+,2})$. The $\xi_{1-4} = Ng_2^2/(c\gamma_4)$ and $\xi_{1-3} = Ng_1^2/(c\gamma_3)$ are respectively the absorption coefficients of the atomic transitions |1>–|4> and |1>–|3>.

Using spatial Fourier transformation $\Psi_\sigma(t,z) = \int_{-\infty}^{\infty} dk e^{ikz} \Psi_{k,\sigma}(t)$ and assuming that initially the field $\Psi_{+,1}(t_o,z)$ is determined by the probe field pulse $A_{+,1}(t_o,z)$ (the other fields are absent: $A_{-,1}(t_o,z) = A_{+,2}(t_o,z) = 0$; $\Omega_{+,2}(t_o) = \Omega_{-,1}(t_o) = 0$), we obtain the following solutions of Eqs. (6) for the Fourier components $\Psi_{k,+2}(t)$ and $\Psi_{k,\pm 1}(t)$:

$$\Psi_{k,+,1}(t) = \frac{\phi(t_o,k)}{\phi(t,k)} \exp\{-i\int_{t_o}^{t} dt' \omega_k(t')\} \Psi_{k,+,1}(t_o), \tag{7a}$$

where

$$\Psi_{k,-,1} = \chi_-(k)\Psi_{k,+,1}, \quad \Psi_{k,+,2} = \chi_+(k)\Psi_{k,+,1}, \tag{7b}$$

the frequency

$$\omega_k(t) = -i\mu_s(t')\{1 + i(k-k_o)/\xi_{+,1} - \phi(t,k)\}/\phi(t,k), \tag{7c}$$

determines the nonstationary dispersion relation for the three color coupled lights, $\chi_-(k) = \frac{\{1+i(k-k_o)/\xi_{+,1}\}}{\{1-i(k+k_o)/\xi_{-,1}\}}$, $\chi_+(k) = \frac{\{1+i(k-k_o)/\xi_{+,1}\}}{\{1+i(k-k_o)/\xi_{+,2}\}}$, and

$$\phi(t,k) = \{\alpha_{-,1}(t)\chi_-(k) + \alpha_{+,1}(t) + \alpha_{+,2}(t)\chi_+(k)\}.$$

We note that the solutions of Eq. (7) actually generalize the results obtained for the two-color stationary light [20,21]. Before detail analysis of Eq. (7), we point out an important universal property of the interaction between the three-light fields: the relations (7b) between the field's Fourier components $\Psi_{k,+2}(t)$ and $\Psi_{k,\pm1}(t)$ are independent of the control laser fields amplitudes and of the decay constant $\gamma_2$. Such relation between the fields indicates a large similarity in their dynamics. Thus the quantum property of the fields including their classical spatial and temporal behavior can be perfectly similar to each other (see below).

The spatial Fourier integrals for the fields $\psi_\sigma(t,z)$ can not be analytically solved at most general conditions of the field interactions owing to the complicated dispersion relation (7c). We, however, solve the integral of $\psi_\sigma(t,z)$ using approximation analysis for some important properties in the coupled light dynamics at some special cases of interactions. Most important properties of coupled lights can be easily found from the dispersion relation (7c) in the limit of an optically dense medium, $k/\xi_\sigma \ll 1; k_o/\xi_\sigma \ll 1$, and weak relaxation between the two lowest levels $\tilde{\gamma}_2 \ll 1$, which usually observed in atomic gases or defected crystals:

$$Lim_{k/\xi\ll 1; k_o/\xi\ll 1} \omega_k(t) \cong$$
$$-i\gamma_2 + \mu_s(t)\{(1-\tilde{\gamma}_2)\alpha_{+,1}(t)(k-k_o)/\xi_{+,1} + \alpha_{+,2}(t)(k-k_o)/\xi_{+,2} - \alpha_{-,1}(t)(k+k_o)/\xi_{-,1}\} - i\delta\omega_{kk}''(t)k^2/2 + \ldots,$$
$$\tag{8}$$

where $\delta\omega_{kk}''$ is the term determined by the main absorption and dispersion effects in evolution of the coupled multi-color light. Here it should be noted that decoherence constant $\gamma_2$ gives the upper limit for the coupled light lifetime in the medium (if the optical density is not so large) [23, 24].

Using Eq. (9) we obtain the group velocity v(t):

$$v(t) = \frac{\partial}{\partial k}\omega_k(t)\Big|_{k=0} \cong \frac{c}{N}\left\{\frac{(1-\tilde{\gamma}_2)\Omega_{+,o1}^2(t)}{g_1^2} + \frac{\Omega_{+,o2}^2(t)}{g_2^2} - \frac{\Omega_{-,o1}^2(t)}{g_1^2}\right\} \quad (9)$$

From Eq. (9), we conclude that all three lights pulses propagate together with the same velocity, which is independent of the spectral detunings $\Delta_{\pm,1}$, $\Delta_{+,2}$, or the atomic relaxation constants $\gamma_{\pm,1(2)}$. Therefore, we can control the coupled group velocity $v(t)$ by manipulating control Rabi frequencies $\Omega_\sigma$, and the velocity direction is either +z or –z axis.

For further analysis of the three-color coupled light we analyze the field amplitude $A_{\pm,m}(t,z)$ for a Gauss shape of the initial probe field, $A_{+,1}(t<t_o,z) = A_{+,o1}\exp\{-\frac{1}{2}(z-v_{+,1}t)^2/l_o^2\}$, where $l_o$ ($A_{+,o1}$) is a longitudinal size (amplitude) of the light pulse. Assuming $k_o = 0$ and no other fields interacted for convenience, $A_{-,1}(t<t_o,z) = A_{+,2}(t<t_o,z) = 0$, the $A_\sigma$ becomes:

$$A_\sigma(t,z) = A_{+;o}\{\Omega_\sigma(t)l_o/[\Omega_{+,1}(t_o)l(t)]\}\exp\{-\tfrac{1}{2}[z-z_\sigma(t)-\int_{t_1}^t v(t')dt' - v_{+,1}(t_1)t_1]^2/l^2(t)\}, \quad (10)$$

where

$$l(t) = \sqrt{[l_o^2 + \int_{t_o}^t dt'\delta\omega_{kk}''(t')]} \quad (11a)$$

is a new pulse length of the coupled light envelope and $z_{\sigma=+,1;+,2;-1}$ are the spatial shifts of the different field shapes:

$$z_{+,1(2)}(t) = [1-\alpha_{+,1(2)}(t)](1/\xi_{+,1(2)}) - \alpha_{+,2(1)}(t)(1/\xi_{+,2(1)}) + \alpha_{-,1}(t)(1/\xi_{-,1}), \quad (11b)$$

$$z_{-,1}(t) = -[1-\alpha_{-,1}(t)](1/\xi_{-,1}) - \alpha_{+,1}(t)(1/\xi_{+,1}) - \alpha_{+,2}(t)(1/\xi_{+,2}). \quad (11c)$$

As seen in Eqs. (10) and (11) the field shapes $A_{+,2}$ and $A_{+,1}$ collide with each other more tightly in space, whereas the field $A_{-,1}$ is shifted farther from these pulses. The distance between these pulses, however, becomes smaller as the medium is optically denser. Thus solution (10) shows a strong coupling of the three light fields in space in an optically dense medium, where they propagate and evolve together in the medium. In fact, the amplitudes of these coupled fields can be manipulated by controlling their Rabi frequencies: $A_\sigma(t,z) \sim \Omega_\sigma(t)A_{+;o}$. Therefore after the photonic trapping through the stationary light it is possible to get only one arbitrary field $A_\sigma$ (if $\Omega_\sigma \neq 0$) from the coupled fields. Some similar properties of light control has been studied recently for a two-color stationary light scheme [20,21], and the quantum nature of the stationary light pulses has been fully analyzed [18,21]. Here in this article we analyze most of the distinctive properties of the three-color stationary light, which are concerned with its spectral properties.

In contrast to the two-color stationary light [20,21] the solution (10) shows that the new quantum field is produced from the stationary light interactions at different frequency: $A_{+,2}$ ($A_{-,1}$) if $\Omega_{+,2} \neq 0$ ($\Omega_{+,1} \neq 0$) and $\Omega_{-,1}=0$. Detail analysis of solution (10) shows that the total energy of the field is conserved, and each field's energy is proportional to $\Omega_{+,2;+1}$.

From Eqs. (9) and (10), stationary light condition (v=0) can be derived:

$$\Omega_{+,o1}^2 / g_1^2 + \Omega_{+,o2}^2 / g_2^2 - \Omega_{-,o1}^2 / g_1^2 = 0, \tag{12}$$

where we took into account a weak relaxation processes between the two lowest levels so $\tilde{\gamma}_2 \ll 1$. Thus satisfying Eq. (12) the two light fields $A_{+,2}(t,z)$ and $A_{+,1}(t,z)$ respectively at frequencies $\omega_{+,2}$ and $\omega_{+,1}$ must be stopped in the medium due to coupled interactions with the field $A_{-,1}(t,z)$ at frequency $\omega_{-,1}$ propagating opposite against the other fields. Using Eq. (12) we can obtain a space relation for the coupled fields: $z_{+,1(2)} = (1/\xi_{+,1(2)})$ and $z_{-,1} = -(1/\xi_{-,1})$. It should be noted that both $z_{+,2}$ and $z_{-,1}$ are overlapped in space for the case of two-color stationary light [20].

Using Eqs. (12), (7c) and (8) we obtain the following second order dispersion term $\delta\omega_{kk}^{"}$:

$$\delta\omega_{kk}^{"} = \frac{c^2}{N^2}\{\frac{\Omega_{+,o1}^2}{g_1^2}[\frac{\gamma_{-,1}+\gamma_{+,1}}{g_1^2}] + \frac{\Omega_{+,o2}^2}{g_2^2}[\frac{\gamma_{-,1}}{g_1^2}+\frac{\gamma_{+,2}}{g_2^2}]\}, \tag{13}$$

which determines the main influence on the spatial spreading of the stationary pulse shape see Eq. (11a). The dispersion (or pulse spreading) is caused by the absorption of the field spectral components which proportional to $\delta\omega_{kk}^{"}k^2/2$ and results in energy loss of the stationary light. The final energy of the light field $E_\sigma$ comparing to the initial energy $W_o$ is: $W_\sigma = \frac{\omega_\sigma}{\omega_{+,1}}(l_o/l(t_1))W_o$. Form Eq. (11a) we already knew that longer interaction time of the stationary light spreads the pulse envelope. Therefore it is important to find out a condition to minimize the light spreading while keeping considerably long interaction time to enhance the nonlinear interaction [17, 20]. This condition is obtained by the frequently adjustment of the light fields:

$$\Delta_{+,2} = -\{\frac{g_2^2}{g_1^2}\Delta_{-,1} + \frac{\Omega_{+,o1}^2}{\Omega_{+,o2}^2}\frac{g_2^4}{g_1^4}(\Delta_{-,1}+\Delta_{+,1})\}. \tag{14}$$

Eq. (14) actually minimizes the dispersion term $\delta\omega_{kk}^{"}$ in Eqs. (11a) and (13). Thus, applying Eq. (14) into Eq. (13) we get:

$$\delta\omega_{kk}^{"} = \{2v_{+,1}\xi_{1-3}^{-1} + v_{+,2}(\xi_{1-3}^{-1}+\xi_{1-4}^{-1})\}, \tag{15}$$

Eq. (14) determines the most stable (optimal) condition in evolution of the spatial shape of the three-color pulses. The stable evolution condition Eq. (14) can be realized in some frequency range for detunings $\Delta_{\pm,1}$ and $\Delta_{+,2}$. There are two different cases of the optimal frequency adjustment:

A) Symmetrical case of the spectral control

$$\Delta_{+,1} = -\Delta_{-,1} \text{ and } \Delta_{+,2} = -(g_2^2/g_1^2)\Delta_{-,1}. \tag{16}$$

Each of these relations also corresponds to the minimum spreading condition of the two-color stationary lights if $\Omega_{+,1} = 0$ or $\Omega_{+,2} = 0$ [20,21]. The minimum spatial spreading condition (14) depend here on the Rabi frequencies $\Omega_{+,om}^2$ (m=1,2) only through the velocities $v_{+,om}$. However we can not equate with zero the Rabi frequency (in particularly $\Omega_{+,1} = 0$) if we want to have a nonzero relevant field component in the stationary light (since $E_{+,m}= 0$, if $\Omega_{+,m} = 0$). Thus the three-color stationary light envelope will spread only two-times faster than that of the two-color stationary light if $\Omega_{+,1} = \Omega_{+,2}$ and $g_2=g_1$ . We also note that the two conditions in Eq. (16) can easily be realized experimentally if the temporal spreading of the stationary light is controlled.

B) The three-color stationary light has an additional function of frequency manipulations. We can vary the frequency detuning in particular $|\Delta_{+,2}|$ in more large frequency range by minimizing the detuning $|\Delta_{+,1}|$ and keeping the condition $\Delta_{-,1} + \Delta_{+,1} = \max \neq 0$. Assuming $(\Omega_{+,o1}^2/\Omega_{+,o2}^2)(g_2^2/g_1^2) \gg 1$ we obtain $\delta\omega_{+,2} \cong -(\Omega_{+,o1}^2/\Omega_{+,o2}^2)(g_2^4/g_1^4)\delta\omega_{-,1}$ and $|\delta\omega_{+,2}|_{g_2/g_1>1} \gg |\delta\omega_{+,1}|$. Such possibility of large frequency detuning $\delta\omega_{+,2}$ looks especially interesting for the spectral control of the resonant interactions between the light and the some selected quantum systems.

### III. DISCUSSIONS AND CONCLUSION

We summarize the followings from Eqs. (7) and (10) for the coupled three-color light:

1) We can control stationary two-color or three-color lights by simple varying the control laser field taking into account the condition of stationary light (see Eq. (12)). Interaction time of the the stationary light can be considerably extended by frequency adjustment of the fields using Eq. (14).

2) The three-color stationary light can switch into one or two traveling waves. For a single traveling wave we can keep the initial frequency $\omega_{+,1}$ or to convert it into $\omega_{+,2}$ or $\omega_{-,1}$: For the case of $\omega_{-,1}$ the field propagates backward. Such wavelength conversion represents a special interest in optical quantum communications. From Eqs. (7a) and (7b) immediately after the switching we get the field $\Psi_{k,m}(t)$:

$$\Psi_{k;\sigma}(t=t_1)\big|_{\Omega_\sigma \neq 0} = \exp\{-i\int_{t_o}^{t_1} dt'\omega_k(t')\}\Psi_{k,+,1}(t_o) \quad (17)$$

For the fast switching processes less than the stationary lifetime, $t_1-t_o$, using Eqs. (9), (11), (13) and (14), we obtain $\omega_k(t_1-t_o) \ll 1$ and $\exp\{-i\int_{t_o}^{t_1} dt'\omega_k(t')\} \cong 1$. Thus, from Eq. (17) $\Psi_{k;\sigma}(t=t_1)\big|_{\Omega_m \neq 0} \cong \Psi_{k,+,1}(t_o)$: The initial quantum field $E_{+,1}$ should convert into the new quantum field $E_{+,2}$ or $E_{-,1}$ without any irreversible energy loss. In particular for the classical Gaussian field we obtain the following from Eq. (10):

$$A_\sigma(t,z) = A_{+;o}\{\Omega_\sigma(t_1)/\Omega_{+,1}(t_o)\}\exp\{-\tfrac{1}{2}[z-\int_{t_1}^{t}v(t')dt'-v_{+,1}(t_1)t_1]^2/l_o^2\} \quad (18)$$

Eq. (18) presents an efficient wavelength conversion of the quantum light pulse through three-color stationary light phenomenon. Such near perfect wavelength conversion should be attractive in many applications. Moreover quantum manipulations of multi-color stationary light will be more interesting in the weak quantum light interactions with a resonant quantum system. Such processes should be potential for enhancement of the nondemolition measurements in nonlinear quantum optics particularly in stronger dipole-dipole interactions of the polariton-associated single-photon field [25, 26].

This work was supported by Korea Research Foundation Grant No. KRF 2003-070-C00024 and the Quantum cryptography project of Korean Ministry of Science and Technology.

FIG. 1

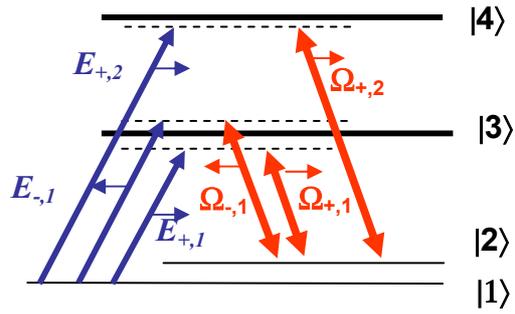

FIG. 1. The diagram of the interaction of the three weak lights $E_{+,2}$, $E_{+,1}$ and $E_{-,1}$ and three control laser fields with their Rabi frequencies $\Omega_{+,2}$, $\Omega_{+,1}$ and $\Omega_{-,1}$ with the four level medium at the condition of two-photon resonance.